\newcommand\ignore[1]{}

\documentclass[a4paper,11pt]{article}
\pdfoutput=1
\usepackage{graphicx}
\usepackage{jheppub}
\usepackage{color}
\usepackage{amssymb}
\usepackage{amsmath}
\usepackage{amssymb}
\usepackage{tocloft}
\usepackage{float}
\usepackage{braket}
\usepackage{url}
\usepackage{subcaption}
\usepackage{upgreek}
\usepackage{outlines}
\usepackage{braket}
\usepackage{appendix}
\usepackage{placeins}
\usepackage{multirow}

\numberwithin{equation}{section}

\def\F{\mathcal{F}}
\def\d{\partial}

\def\a{\alpha}

\def\eps{{\varepsilon}}
\usepackage{bm}

\def\n{\nu}

\def\G{\Gamma}
\def\g{\gamma}
\def\k{\kappa}
\usepackage{caption}
\usepackage{subcaption}

\def\t{\tau}
\def\tq{\tilde{q}}
\def\Om{\Omega}
\def\p{\rho}

\def\th{\theta}

\def\onec{\left(\begin{array}{c}}
\def\cend{\end{array}\right)}
\def\cc{\left(\begin{array}{cc}}
\def\ccend{\end{array}\right)}
\def\ccc{\left(\begin{array}{ccc}}

\def\cccend{\end{array}\right)}
\def\cccc{\left(\begin{array}{cccc}}

\def\p{\psi}

\def\ccccend{\end{array}\right)}

\def\t{\tau}

\newcommand{\calN}{\mathcal{N}}

\def\hi{\hat{I}}
\renewcommand\Re{\operatorname{Re}}
\renewcommand\Im{\operatorname{Im}}

\newcommand{\labeleq}[1]{\label{eq:#1}}

\renewcommand\mod[1]{\text{ mod }#1}

\newcommand{\eq}[1]{\begin{align}#1\end{align}}
\newcommand{\seq}[1]{\begin{align*}#1\end{align*}}

\newcommand{\subeqslabel}[2]{\begin{subequations}\begin{align}#1\end{align}\label{#2}\end{subequations}}
\def\de{\delta}
\def\De{\Delta}

\def\tc{\tilde{C}}

\def\Z{\ensuremath{\mathbb{Z}}}
\def\R{\ensuremath{\mathbb{R}}}

\usepackage[bottom]{footmisc}
\def\C{\ensuremath{\mathbb{C}}}

\def\hg{\hat{g}}

\def\gcd{\operatorname{gcd}}
\newcommand{\SL}{\operatorname{SL}}
\newcommand{\sltz}{\SL(2,\Z)}
\newcommand{\sltr}{\SL(2,\R)}
\newcommand{\smallmat}[1]{\left(\begin{smallmatrix}#1\end{smallmatrix}\right)}
\def\kl{\operatorname{Kl}}

\def\dnp{d_n^{(p)}}
\def\hsum{\sum_{\substack{0\le h<k\\ \gcd(h,k)=1}}}
\def\sumh{\hsum}
\def\abcd{\smallmat{a&b\\c&d}}

\def\Q{\mathbb{Q}}

\def\hQ{\hat{\Q}}

\newcommand{\ghk}{\g_{k,h}}
\newcommand{\gkh}{\ghk}
\newcommand{\gpkh}{\g_{k,h}^{(p)}}

\newcommand{\gkhp}{\gpkh}

\renewcommand{\Z}{\mathbb{Z}}

\def\calN{\mathcal{N}}

\def\fp{F^{(p)}}
\def\ggg{\Gamma^\infty\backslash\G/\G^\infty}
\def\ggpg{\Gamma^\infty\backslash\G_0(p)+/\G^\infty}
\def\zpt{Z_p(\t)}
\def\gop{\G_0(p)}
\def\gopp{\G_0(p)+}

\newcommand\numberthis{\addtocounter{equation}{1}\tag{\theequation}}

\title{Exact Half-BPS Black Hole Entropies in CHL Models from Rademacher Series}

\author[a]{Richard Nally}
\affiliation[a]{Department of Physics, Stanford University}
\emailAdd{rnally@stanford.edu}

\begin{document}

\abstract{The microscopic spectrum of half-BPS excitations in toroidally compactified heterotic string theory has been computed exactly through the use of results from analytic number theory. Recently, similar quantities have been understood macroscopically by evaluating the gravitational path integral on the M-theory lift of the AdS$_2$ near-horizon geometry of the corresponding black hole. In this paper, we generalize these results to a subset of the CHL models, which include the standard compactification of IIA on $K3\times{T}^2$ as a special case. We begin by developing a Rademacher-like expansion for the Fourier coefficients of the partition functions for these theories, which are modular forms for congruence subgroups. We then interpret these results in a macroscopic setting by evaluating the path integral for the reduced-rank $\calN=4$ supergravities described by these CFTs.}

\maketitle

\section{Introduction}
\label{sec:intro}

Amongst the foremost successes of string theory as a framework for quantum gravity is its ability to provide a microscopic interpretation of the Bekenstein-Hawking entropy \cite{StromingerVafa}. Strominger and Vafa found that the area law entropy emerges naturally from a saddle-point evaluation of the elliptic genus of the worldsheet theory describing a certain class of BPS black holes. In the twenty years since the original Strominger-Vafa calculation, this strategy of studying BPS black hole entropy by matching micro- and macroscopic calculations of supersymmetric indices has been the subject of significant study. 

A particularly fruitful example of this program has been the study of half-BPS black holes in heterotic string theory compactified to four dimensions on $T^6$. These black holes are ``small", in the sense that their horizons have vanishing area in the strict $\a'=0$ limit, and have been studied extensively \cite{0409,0502,0507,Dabholkar:2004dq}. Their entropies can be derived from the study of worldsheet degeneracies enumerated by a supersymmetric index of the form\footnote{As defined here, this counting function enumerates short BPS multiplets, rather than the number of states directly. These two counts are simply related by a constant of proportionality, which for brevity we will omit.} \eq{Z(\t) = \frac{1}{\De(\t)},} where $\t$ is the modular parameter of the worldsheet torus, and \eq{\De(\t) = q\prod_{n=1}^\infty\left(1-q^n\right)^{24} \equiv \eta^{24}(\t)\label{eq:DeDef}} is the modular discriminant \cite{0409}, where as usual we have defined \eq{q = e^{2\pi i \t} \label{eq:qDef}.} 

Crucial to the microscopic study of these black holes are the modular properties of this counting function\footnote{For a concise introduction to modular forms, see e.g. \cite{123}.}. In particular, the partition function $Z(\t)$ is a weight $-12$ modular form for the modular group $\sltz$, meaning it obeys the functional equation \eq{Z\left(\frac{a\t+b}{c\t+d}\right) &= (c\t+d)^{-12}Z(\t)} for $\left(\begin{smallmatrix}a&b\\c&d\end{smallmatrix}\right) \in \sltz$. From the $\sltz$ matrix $\smallmat{1&1\\0&1}$, we see that $Z(\t+1) = Z(\t)$, so that $Z(\t)$ has a Fourier expansion \eq{Z(\t) = \sum_{n=-1}^\infty d_nq^n.} That the leading term is $q^{-1}$ follows directly from Eq. \ref{eq:DeDef}. The coefficients $d_n$ are the degeneracy of BPS states at energy level $n$, so the problem of computing BPS black hole entropy $S_n \equiv \ln{d_n}$ thus reduces to the computation of the Fourier coefficients $d_n$.

An exact series expression for the $d_n$ was proven by Rademacher in the early twentieth century \cite{RademacherOriginal,Rademacher43}. Elementary proofs are given in standard analytic number theory texts, such as \cite{apostol,RademacherBook}, as well as in \cite{FareyTail}; more detailed discussion can be found in \cite{ChengRademacher}. The coefficients are given by \cite{RademacherOriginal,Rademacher43,apostol,RademacherBook,FareyTail,ChengRademacher,0502,0507} \eq{d_n = \sum_{k=1}^\infty k^{-14} \kl(n,-1,k) \hi_{13}\left(\frac{4\pi}{k}\sqrt{n}\right), \label{eq:DHrad}} where we have defined the ``Kloosterman sum" \eq{\kl(n,m,k) = \sum_{\substack{0<h<k\\\gcd(h,k)=1}} \exp\left[2\pi i \left(\frac{hn}{k} + \frac{h^{-1}m}{k}\right)\right],\label{eq:KlDef}} where the inverse $h^{-1}$ of $h$ mod $k$ is defined by the condition $hh^{-1}\equiv1\mod{k}$, and \eq{\hi_\n(z) \equiv -i\left(2\pi\right)^\n \int_{\eps-i\infty}^{\eps+i\infty} dt\ t^{-\n-1} e^{t+z^2/4t} = 2\pi\left(\frac{z}{4\pi}\right)^{-\n}I_\n(z)\labeleq{BesselDef}} is (proportional to) a Bessel function. This expansion was first introduced to the string theory literature in \cite{FareyTail}, and applied to the study of half-BPS black holes in $\calN=4$ compactifications in \cite{0502,0507}. For large $z$, we have \eq{\hi_{\n}(z)\sim \frac{e^z}{\sqrt{2}} \left(\frac{z}{4\pi}\right)^{-\n-\frac{1}{2}} \left[1-\frac{4\n^2-1}{8z} + \frac{(4\n^2-1)(4\n^2-9)}{2(8z)^2} - \cdots\right],\label{eq:bessel}} so as the energy level $n$ grows, each term in Eq. \ref{eq:DHrad} is exponentially suppressed relative to the preceding term. At large $n$, we therefore have \eq{d_n \sim \hi_{13}\left(\sqrt{n}\right),\label{eq:leading}} which encapsulates all perturbative corrections\footnote{By ``perturbative" here we mean suppressed by powers of $n$; this is in contrast to the $k>1$ terms, which are exponentially suppressed in $n$. From a gravitational perspective, $n$ is related to the electric and magnetic charges of the black hole, so this is a standard expansion in the charges.} to the black hole entropy; the subleading terms in Eq. \ref{eq:DHrad} similarly correspond to nonperturbative corrections.  

From the microscopic side, we therefore have an exact understanding of all perturbative and nonperturbative corrections to the degeneracy, and therefore the entropy, of the black hole. What remains to be done is to understand these corrections macroscopically. Although the black holes dual to these half-BPS states have classically vanishing horizon area, and therefore are in principle unsuited to a geometric description, it has been shown that $\alpha'$ corrections to the gravitational action give a horizon of finite size in string units \cite{Dabholkar:2004dq}. This allows us to obtain a geometric interpretation of the states, and in particular lets us consider the familiar $AdS_2\times{S}^2$ near-horizon geometry usually associated to extremal black holes. For more details, see e.g. \cite{SenNotes,DabNotes} and the references therein.

Macroscopically, the perturbative corrections to the degeneracy of BPS microstates, i.e. the $k=1$ term in Eq. \ref{eq:DHrad}, were already understood in \cite{0409,0502,0507} using an OSV-like formalism. However, understanding the nonperturbative corrections is more subtle. Intuitively, the correct macroscopic quantity to compute is the gravitational path integral evaluated on the near horizon geometry. The terms in the sum in Eq. \ref{eq:DHrad} are in one-to-one correspondence with the well-known $\sltz$ saddle points of $AdS_3$ gravity, so the Rademacher series should be reproduced exactly by the saddle point expansion of this path integral. To obtain $AdS_3$ saddle points from an $AdS_2$ near-horizon geometry, we consider the M-theory lift of the dual IIA geometry \cite{Murthy:2009dq,Beasley:2006us}. This basic intuition was first put forth in \cite{FareyTail}, and later expanded on in e.g. \cite{deBoer:2006vg,Manschot:2007ha,Murthy:2009dq}.

Making this intuition rigorous requires the use of supersymmetric localization techniques to facilitate the evaluation of the path integral \cite{Dabholkar:2010uh,Dabholkar:2011ec,Gomes:2011zzc,Dabholkar:2014ema}. After localization, the path integral is given exactly by the sum of an appropriate action evaluated on the $\sltz$ saddles; as argued above, this reproduces the structure of the Rademacher series with great precision. Indeed, by evaluating the action, one finds agreement with both the Bessel functions and phases in Eq. \ref{eq:DHrad}. Roughly, the Bessels are given by the local part of the action, and the Kloosterman sums stem from topological terms \cite{Dabholkar:2014ema}. In this manner, the macroscopic gravitational physics can be seen to reproduce the nonperturbatively exact microscopic results.

In this paper, our goal is to generalize this analysis to a class of heterotic orbifolds known as the CHL models \cite{CHLoriginal,Chaudhuri:1995bf,Chaudhuri:1995dj}. These models describe heterotic string theory compactified on orbifolds of $T^6$, or dually type II compactified on orbifolds of $K3\times T^2$. These orbifolds preserve supersymmetry, so on the gravity side these theories have $\calN=4$ supersymmetry in four dimensions. In general, therefore, these theories can admit half- and quarter-BPS black hole solutions. The counting of black hole states in these theories was initiated in \cite{0502,0507}, and was advanced significantly with the introduction of explicit forms for the counting functions in \cite{JatkarSen}. This problem is additionally discussed in e.g. \cite{David:2006yn,CHLcomposite,Govindarajan:2010fu,Gomes:2015xcf}.

The set of all discrete symmetries of the K3 sigma model, and therefore the set of CHL models, was recently enumerated \cite{Persson:2015jka,Paquette:2017gmb,Cheng:2016org,Gaberdiel:2011fg}. From these results, as well as the earlier results of e.g. \cite{JatkarSen,CHLcomposite}, we see that the twisted-sector half-BPS counting functions are $\eta$-quotients, i.e. functions of the form \eq{Z(\t) = \frac{1}{\prod_{a\in\Z}\eta(a\t)^{m(a)},}} where the powers $m(a)$ are always integral but not necessarily positive. Whereas the counting function $Z(\t) = \frac{1}{\De(\t)}$ discussed above is modular for $\sltz$, these functions are not.  Because of this, the results of Eq. \ref{eq:DHrad} do not apply to these counting functions, and a generalization of the classical Rademacher series is needed to precisely study these theories. However, they are modular for other discrete subgroups of $\sltr$, known as Atkin-Lehner groups, which will provide a powerful tool with which we will be able to study these functions.

In this paper, we will use the results of \cite{Duncan:2009sq,ChengRademacher} to develop a generalized Rademacher expansion to study the degeneracies of BPS black holes in CHL models precisely, and compare this microscopic result to a detailed macroscopic calculation. Throughout, we will consider as a benchmark a particularly simple subset of the CHL models, called the prime-$p$ models \cite{JatkarSen}. The $p=1$ model is simply the standard $T^6$ theory, so our benchmark models encapsulate the results of \cite{0409,0502,0507,Dabholkar:2014ema}. In these theories, the appropriate half-BPS counting functions are  \cite{JatkarSen,David:2006yn,Gomes:2015xcf,Govindarajan:2010fu,Persson:2015jka,Paquette:2017gmb} \eq{Z_p(\t) = \frac{1}{\left[\eta(\t)\eta(p\t)\right]^w},} where $p\in\{1,2,3,5,7,11\}$ and \eq{w = \frac{24}{1+p}.} These functions admit Fourier series of the form \eq{Z_p(\t) = \sum_{n=-1}^\infty \dnp q^n.} We will show that the Fourier coefficients $\dnp$ are given by \seq{\dnp = &\sum_{\substack{k>0\\\gcd(p,k)=p}} \sumh C_p(k,h) \exp\left[2\pi i\left(n\frac{h}{k} - \frac{h^{-1}}{k}\right)\right]\hi_{1+w}\left[\frac{4\pi}{k}\sqrt{n}\right] \\ + &\sum_{\substack{k>0\\\gcd(p,k)=1}} \sumh C_p(k,h) \exp\left[2\pi i\left(n\frac{h}{k} - \frac{(ph)^{-1}}{k}\right)\right] \hi_{1+w}\left[\frac{4\pi}{k}\sqrt{\frac{n}{p}}\right], \numberthis \label{eq:IntroRad}} where $C_p(k)$ is a prefactor. This equation is an exact, convergent series expansion for the degeneracies of half-BPS excitations of any energy level, and is our main result.  In the course of this paper, we will derive this result from a microscopic perspective, and then reinterpret it macroscopically.

The outline of this paper is as follows. In Section \ref{sec:CHL}, we will introduce CHL models and their counting functions, and derive Eq. \ref{eq:IntroRad}.  Next, in Section \ref{sec:macro}, we will generalize the Farey tail for $1/\De$ to include black holes for CHL models, and in so doing will reproduce Eq. \ref{eq:IntroRad} macroscopically. Finally in Section \ref{sec:conclusion} we will conclude with a brief discussion and outlook. 

\section{CHL Models and Their Counting Functions}
\label{sec:CHL}

In this section we will introduce the CHL models and their counting functions. We begin in Section \ref{sec:CHLconstruction} with a review of the details of these models. In Section \ref{sec:functs} we will introduce the twisted sector half-BPS counting functions $Z_p(\t)$. Finally, in Section \ref{sec:CHLrad} we will introduce Rademacher series for the $Z_p(\t)$, and derive Eq. \ref{eq:IntroRad}.

\subsection{The Prime-$p$ CHL Models}
\label{sec:CHLconstruction}
In this section we will briefly outline the construction of the CHL models. Although there exists an equivalent definition in terms of heterotic compactifications, for these purposes it is most convenient to work in the Type II picture, where these models are constructed in terms of an orbifold of the non-linear sigma model (NLSM) on $K3$, compactified down to four dimensions from six on a $T^2$. The first CHL models were constructed in \cite{CHLoriginal,Chaudhuri:1995bf,Chaudhuri:1995dj}; more recently, this construction has been generalized, and all possible CHL models have been classified \cite{Paquette:2017gmb,Gaberdiel:2011fg,Cheng:2016org}. Here we will simply sketch the construction; the interested reader is referred to \cite{Paquette:2017gmb} for more details on the construction of these models.

The orbifold theory is constructed from a symmetry $g\in{O}\left(\G^{4,20}\right)$ of the K3 NLSM of order $N$, as well as an order-$N$ translation\footnote{A more general class of CHL models can be constructed by relaxing the condition that the orders of $g$ and $\de$ be equal. For our purposes, this more general construction, discussed at length in \cite{Paquette:2017gmb}, will not be needed, and it will be sufficient to demand that the two orders be equal.} $\de$ along a cycle $S^1\subset T^2$. Early constructions of CHL models focused on the case when the symmetry $g$ of the K3 NLSM is a genuine geometric symmetry of the K3 surface; however, now there are known examples of CHL models where $g$ has no geometric interpretation. Pairs $\hg \equiv (g,\de)$ generating inequivalent CHL models have been shown to be related to (equivalence classes of) elements of the Conway group $Co_0$ \cite{Persson:2015jka,Persson:2017lkn,Paquette:2017gmb,Gaberdiel:2011fg,Cheng:2016org}; we will for convenience denote the $Co_0$ equivalence class dual to $\hg$ simply by $g$.

A tremendous amount of information about the CHL model generated by $\hg$ can be gleaned from the group theoretic properties of the dual Conway element. Amongst the most important data to be gleaned from the Conway description is the ``frame shape" of the symmetry, which summarizes the eigenvalues of the K3 NLSM symmetry $g$ in the 24 dimensional representation of ${O}\left(\G^{4,20}\right)$. The frame shape $\pi_g$ associated to $g$ is a formal product \eq{\pi_g = \prod_{a|N} a^{m(a)},} where \eq{\sum_{a|N} am(a)=24.} Each element has 24 eigenvalues $\lambda_i$, all of which live on the unit circle, given by the solutions to the characteristic polynomial \eq{\det\left(\lambda\mathbb{I}-g\right) = \prod_{a|N}\left(\lambda^a-1\right)^{m(a)} = 0.}  These frame shapes will be extremely relevant to both micro- and macroscopic counting of BPS states. 

In this paper we will be primarily interested in a particularly simple subset of CHL models known as the prime-$p$ CHL models, so named because the order $p$ of the orbifold is prime\footnote{In a slight abuse of notation, we will consider the unorbifolded compactification to be a prime-$p$ model with $p=1$.}. These models are defined for\footnote{From a strictly number theoretic perspective, it appears that we should allow $p=23$ as well. The frameshape $\pi_{23} = 1^123^1$ has strictly integral powers, and similarly the eta-product $\eta_{23}(\t) = \eta(\t)\eta(23\t)$ is a well-defined weight-1 cusp form for $\G_1(23)$. However, from the classification in \cite{Persson:2015jka,Paquette:2017gmb,Cheng:2016org,Gaberdiel:2011fg}, it is clear that there does not exist an order-23 symmetry of the K3 NLSM, and correspondingly there is no CHL model with $N=23$. Thus, although from the results of this section we could compute the Fourier coefficients of $Z_{23}(\t) = 1/\eta_{23}(\t)$ via a Rademacher series, doing so would not count states in a string theory. We thank J. Duncan for emphasizing this point.}  \eq{p=1, 2, 3, 5, 7,\text{ and } 11 \label{eq:Pchoices},} and characterized by a frameshape of the form \eq{\pi_p = 1^w p^w,\label{eq:pip}} where \eq{w = \frac{24}{p+1}.\label{eq:wdef}} Some important information about these models is collected in Table \ref{tab:list}.  For $p\neq11$, they are constructed from geometric symmetries of the K3 surfaces, known as ``Nikulin automorphisms," which preserve the K3 holomorphic two-form \cite{1997alg.geom..1011N}. In general, these automorphisms have fixed points; the accompanying shift along $S^1\subset T^2$ renders the orbifold freely acting, and we end up on a six-manifold with the local structure of $K3\times T^2$. 

\begin{table}
\begin{centering}
\begin{tabular}{|c|c|c|c|c|}\hline
$p$ & $\pi_p$ &  $w$ & $\psi_p$ & $r_p$ \\\hline
1 & $1^24$ & 12 & 1 & 28\\\hline
2 & $1^82^8$ & 8 & 1 & 20\\\hline
3 & $1^63^6$ & 6 & 1 & 16\\\hline
5 & $1^45^4$ & 4 & 1 & 12\\\hline
7 & $1^37^3$ & 3 & $\left(\frac{-7}{d}\right)$ & 10\\\hline
11 & $1^211^2$ & 2 & 1 & 8 \\\hline

\end{tabular}
\caption{The prime-$p$ CHL models. For each model, we give the order $p$ of the orbifold, the frameshape $\pi_p$ of the model, and the weight $w$ and multiplier system $\psi_p$ of $\eta_g$ under $\G_0(N)$ transformations. The multiplier systems are given in terms of the Jacobi symbols defined in Eq. \ref{eq:JacobiSymbol}. For later reference, we also give the rank $r_p$ of the gauge group, computed using Eq. \ref{eq:pRank}. The information contained in this table is adapted from \cite{Paquette:2017gmb,He:2013lha,Govindarajan:2010fu,Bossard:2017wum,Cheng:2016org}.}
\label{tab:list}
\end{centering}
\end{table}

 The usual compactification of IIA on $K3\times T^2$, which is the $p=1$ CHL model, has a gauge group of rank 28, and a moduli space of the form \eq{\mathcal{M} = \frac{O(6,22)}{O(6)\times O(22)} \times \frac{\sltr}{U(1)},} quotiented out by a discrete duality group.  Other CHL models have gauge group of reduced rank $r_g$.  In analogy with the usual case of IIA on $K3\times T^2$, the moduli space $\mathcal{M}(\hg)$ of the CHL model associated to $\hg$ is locally \eq{\mathcal{M}(\hg) = \frac{O(6,r_g-6)}{O(6)\times O(r_g-6)} \times \frac{\sltr}{U(1)}.} Globally, the moduli space is quotiented by the action of a discrete duality group. For the prime-$p$ models, the rank $r_p$ of the gauge group is given explicitly by \cite{Bossard:2017wum,JatkarSen} \eq{r_p = \frac{48}{p+1} + 4.\label{eq:pRank}}

 At generic points in moduli space, the gauge group is $U(1)^{r_p}$. States in the CHL models have $r_p$-dimensional electric and magnetic charge vectors $Q$ and $P$, respectively. From these vectors, we can construct three quadratic invariants: $Q^2$, $P^2$, and $Q \cdot P$. In the unorbifolded case, these invariants are all quantized in integral units. However, for general CHL models, we have a more complicated quantization given by \cite{Paquette:2017gmb,JatkarSen},   \eq{\frac{Q^2}{2} \in \frac{1}{N}\Z, \ \ \frac{P^2}{2} \in \Z, \ \  Q\cdot P \in \Z.\label{eq:pquant}}

The duality groups of CHL models are in general rather complicated \cite{Persson:2015jka,Persson:2017lkn,Paquette:2017gmb}. For our purposes, the most interesting component of the full duality group will be the $S$-type dualities. The first class of $S$-duality transformations are given by a $\G_1(N)$ action on the heterotic axiodilation $S$ and the charges $Q$ and $P$, i.e. \eq{S\to \frac{aS+b}{cS+d}, \ \ \left(\begin{array}{c}Q\\P\end{array}\right) \to \left(\begin{array}{cc} a&b \\ c&d \end{array}\right) \left(\begin{array}{c}Q\\P\end{array}\right) \text{ for } \left(\begin{array}{cc}a&b\\c&d \end{array}\right) \in \G_1(N).} This can be extended to a larger class of $S$-dualities that act as $\G_0(p)$ elements on the axiodilation \cite{Paquette:2017gmb}. More recently, it was shown that the $S$-duality groups also contain Fricke involutions \cite{Persson:2015jka,Persson:2017lkn}. These Fricke-type dualities act on the axiodilaton as $S\to -1/NS$, but in general map different CHL models into each other. Fortunately, the prime-$p$ models are self-dual under Fricke duality \cite{Persson:2015jka,Persson:2017lkn}, so we can ignore this subtlety.

\subsection{Half-BPS Counting Functions}
\label{sec:functs}
We will now describe the counting of a class of half-BPS states in these models. First we will construct an important class of modular forms that will be essential to what follows. To each prime-$p$ CHL model, we can associate a modular form $\eta_p(\t)$ defined by\footnote{This construction applies to all CHL models. In particular, to a frameshape $\pi_g = \prod a^{m(a)}$, we can associate a modular form $\eta_g(\t) \equiv \prod \eta(a\t)^{m(a)}.$ In general, this will be a weight $\frac{1}{2}\sum am(a)$ modular form, with multiplier system, for $\G_0(N)$.} \eq{\eta_p(\t) = \eta(\t)^w\eta(p\t)^w.} These functions are weight-$w$ modular forms without multipler system for the congruence subgroup $\G_1(p)\subset\sltz$, defined as  \eq{\G_1(p) \equiv \left\{\left.\abcd\in\sltz\ \right|\ a,c \equiv 1\mod{p}, c \equiv 0\mod{p}\right\}.} The weight $w$ is defined in Eq. \ref{eq:wdef}. These functions extend to modular forms for the larger congruence subgroup $\G_0(p)\supset\G_1(p)$ defined by \eq{\G_0(p) \equiv \left\{\left.\abcd\in\sltz\ \right|\ c \equiv 0\mod{p}\right\},} but in general have nontrivial multiplier systems under this larger group. Notably, the modular discriminant $\De(\t)$ is the modular form associated to the frameshape corresponding to the $p=1$ model, i.e. the usual compactification of heterotic strings on $T^6$.

For $p\neq7$, the $\eta_p$ have trivial multiplier system under all $\G_0(p)$ transformations. However, for $p=7$ the multiplier system is more subtle, essentially because $\eta_7(\t)$ has odd weight. The multiplier of $\eta_7(\t)$ under an element $\g=\smallmat{a&b\\c&d}\in\gop$ is related to whether or not the bottom-right entry $d$ is a quadratic residue with modulo 7.  For $b$ prime and $a\in\Z$, we define the Legendre symbol $\left(\frac{a}{b}\right)$ by \cite{He:2013lha} \eq{\left(\frac{a}{b}\right) = \left\{ \begin{array}{cc} 0 & a\equiv0\mod{b} \\ 1 & a \not\equiv 0\mod{b}\text{ and } \exists x\in\Z \text{ s.t. } a \equiv x^2\mod{p} \\ -1 & a \not\equiv 0\mod{b}\text{ and } \nexists x\in\Z \text{ s.t. } a \equiv x^2\mod{p} \end{array} \right..} For $n\in\Z$ odd and positive with prime factorization $n = p_1^{c_1} p_2^{c_2}\cdots p_m^{c_m}$, we define the Jacobi symbol $\left(\frac{a}{n}\right)$ by \cite{He:2013lha} \eq{\left(\frac{a}{n}\right) = \left(\frac{a}{p_1}\right)^{c_1} \left(\frac{a}{p_2}\right)^{c_2} \cdots \left(\frac{a}{p_m}\right)^{c_m},\label{eq:JacobiSymbol}} where $\left(\frac{a}{p_i}\right)$ is a Legendre symbol. For $\g = \smallmat{a&b\\c&d}\in\G_0(p)$, we then have \eq{\eta_p(\g\cdot\t) = (c\t+d)^w \psi_p(\g) \eta_p(\t),} where as usual by $\g\cdot\t$ we mean $\frac{a\t+b}{c\t+d}$ and we have defined multiplier systems $\psi_p(\g)$ by \cite{He:2013lha} \eq{\psi_p(\g) = \left\{\begin{array}{cc} 1 & p \neq7\\ \left(\frac{-7}{d}\right) & p=7, \ d\text{ odd}\\ \left(\frac{d}{7}\right) & p=7,\ d\text{ even} \end{array}\right.. \label{eq:G0pmult}}  

Because it will be relevant to the discussion in Section \ref{sec:CHLrad}, we will pause now to discuss the cusp structures of the various modular groups considered here. Consider a discrete subgroup $\G\subset\sltr$ commensurable with $\sltz$. Define the ``cusps" of $\G$ as the equivalence classes of the set $\hQ \equiv \Q\cup i\infty$ under the action of $\G$.  We call a modular form for $\G$ a cusp form for $\G$ if it vanishes at all of the cusps of $\G$.

As an example, consider the usual modular group $\sltz$. For any rational number $\frac{h}{k}$, there exists a family $\g_{k,h}$ of $\sltz$ elements that maps the point $\t=i\infty$ into $\frac{h}{k}$. Therefore, it is said that $\sltz$ has only one cusp up to equivalence. From the definition given in Eq. \ref{eq:DeDef}, it is clear that $\De(\t)$ vanishes in the limit $\t\to{i\infty}$. By modular invariance, it also therefore vanishes at all rational points, so $\De(\t)$ is a cusp form for $\sltz$. 

Similarly, the $\eta_p$ are cusp forms with multiplier system for $\gop$, although for these groups the cusp structure is more complicated. The cusp forms $\eta_p(\t)$ still vanish at all rational points, but for $p\neq1$ these points are not all related by $\G_0(p)$ transformations. Thus these subgroups have multiple inequivalent cusps. In particular, for $p\neq1$, $\gop$ always has exactly two cusps: $\t=0$ and $\t=i\infty$. These two points are related by the $\sltz$ transformation $\t\to-1/\t$, but for $p\neq1$ this is not an element of $\gop$. For composite $N$, $\G_0(N)$ will in general have more than two cusps. For instance, the group $\G_0(4)$ has three cusps: $\t=i\infty$, $\t=\frac{1}{2}$, and $\t=0$. 

For our purposes we will need to consider the transformation properties of the $\eta_p$ under not only subgroups of $\sltz$, but also under discrete subgroups of $\sltr$. Consider the ``Fricke involution" \eq{\t\to-\frac{1}{p\t}.} This transformation cannot be written as an element of $\sltz$, but can be embedded in $\sltr$ as the matrix \eq{F_p = \frac{1}{\sqrt{p}} \begin{pmatrix} 0&-1\\p&0\end{pmatrix} \in \sltr.} It is straightforward to verify that $\eta_p$ transforms as a weight-$w$ modular form, in general with multiplier system, under the Fricke involution, so these functions are modular for at least some elements of $\sltr$ not contained in $\sltz$. The modularity of $\eta_p$ extends to a class of $\sltr$ elements known as ``Atkin-Lehner (AL) transformations," \cite{Persson:2015jka,Persson:2017lkn,Paquette:2017gmb,Duncan:2009sq,Duncan:2014vfa} which we will now discuss. 

Consider an integer $N$. Define an ``exact divisor" $e$ of $N$ (written as $e||N$) by the condition \eq{e||N \iff e|N \text{ and } \gcd(e,N/e) = 1.} To each exact divisor $e$ of $N$, we define a family of $\sltr$ elements $W_e(a,b,c,d)$ defined by \eq{W_e(a,b,c,d) = \frac{1}{\sqrt{e}} \begin{pmatrix} ae & b\\ cN&de \end{pmatrix}.\label{eq:WeForm}} Since this matrix must have determinant one, we demand \eq{ade^2 - bcN = e\label{eq:ALdet}.} These are the AL transformations. For $e=1$, we simply obtain $\G_0(N)$. Additionally, since $N||N$, there is always at least one other class of AL transformations, with $e=N$. These are of the form \eq{W_N(a,b,c,d) = \frac{1}{\sqrt{N}}\cc aN&b\\cN&dN \ccend} with \eq{adN - bc = 1.} Taking $a=d=0$ and $b=-1$, we obtain the Fricke involution. 

The only exact divisors of a prime are one and itself, so for the prime-$p$ models we need only consider $\G_0(p)$ and AL transformations of the form $W_p(a,b,c,d)$. It is conventional to package these transformations into the discrete subgroup $\gopp\subset\sltr$, defined by \eq{\gopp \equiv \gop \cup \left\{W_p(a,b,c,d) \big| adp - bc = 1 \right\}.\label{eq:WNForm}} This group is the normalizer of $\gop$ in $\sltr$. Importantly, the groups $\gopp$ have exactly one cusp. This is easy to see, as the Fricke involution $\t\to-1/p\t$ directly maps the cusps $\t=0$ and $\t=-\infty$ into each other.

In general, the $\eta_p(\t)$ have nontrivial multiplier systems under AL transformations. It is straightforward to verify that, for $p\neq7$, the $\eta_p$ transform under Fricke involution as \eq{\eta_p\left(-1/p\t\right) = e^{2\pi i w/4}\left(\sqrt{p}\t\right)^w\eta_p(\t),} i.e. the Fricke involution has multiplier system $e^{2\pi i w/4}$. More generally, for $W_p(a,b,c,d)\in\gopp$ of AL type, the multiplier system is \eq{\psi_p(W_p) = \left\{\begin{array}{cc} e^{2\pi i w/4} & p \neq 7\\ i \left(\frac{c}{7}\right) & p = 7 \end{array}\right.,\label{eq:ALmult}} where the form of $W_p$ is as in Eqs. \ref{eq:WeForm} and \ref{eq:WNForm} and $\left(\frac{c}{7}\right)$ is a Legendre symbol.

We will now use the $\eta_p(\t)$ to count twisted-sector half-BPS states, but first we must define the class of states we will be counting. In the remainder of the paper, we will be concerned with purely electrically charged states half-BPS states. These states have $P^2 = Q\cdot P = 0$, with $Q^2$ nonzero. From Eq. \ref{eq:pquant}, we therefore have \eq{\frac{Q^2}{2} = \frac{n}{p}.\label{eq:Q2p}} In the unorbifolded theory, all half-BPS states are of this form. However, in the orbifold theory, this is no longer true, and so we will only count a subset of twisted-sector half-BPS states. As a further complication, although in the unorbifolded theory the degeneracies of BPS states depend only on these invariants, in the CHL  models in general the denegeracies depend on additional discrete invariants \cite{Paquette:2017gmb,Banerjee:2008pv}.

For large values of $Q^2$, the dependence on these discrete invariants is expected to become unimportant \cite{Paquette:2017gmb}. We will therefore ignore this subtlety, and simply count the twisted-sector analogs of the Dabholar-Harvey states counted in the unorbifolded theory in \cite{0409,0502,0507}. In the heterotic description, these are perturbative states with momentum and winding along a circle in the compact space.  In the unorbifolded theory, any perturbative state with $m$ units of momentum and $w$ units of winding can be dualized into a state with $mw$ units of momentum and one unit of winding. However, in the orbifolded theories, this is no longer true. In particular, the residue $r$ of the winding $w$ modulo the order $p$ of the orbifold determines which twisted sector of the orbifold theory the state lives in. We are interested in twisted sector states, so we take $w\equiv1\mod{p}$. For simplicity, we will take $w=1$. In light of the charge quantization condition in Eq. \ref{eq:Q2p}, were are therefore led to consider fundamental strings with one unit of winding and momentum $n/p$ along $S^1\subset T^6$.  These states can be counted by a simple oscillator count in the unorbifolded theory \cite{Paquette:2017gmb,Govindarajan:2010fu}. 

These counts are well known in the literature. We define a set of counting functions $Z_p(\t)$ which generate the degeneracies $\dnp$ of twisted-sector DH states with $Q^2 = 2n/p$ as \eq{Z_p(\t) \equiv \sum_{n=-1}^\infty \dnp q^n,} where as usual $q=e^{2\pi i\t}$. For the prime-$p$ models, these have been computed explicitly, and are given by \cite{David:2006yn,CHLcomposite,Govindarajan:2010fu,Gomes:2015xcf,Persson:2015jka,Paquette:2017gmb,JatkarSen,Bossard:2017wum,CHLcomposite,He:2013lha} \eq{Z_p(\t) = \frac{1}{\eta_p(\t)} = \frac{1}{\left[\eta(\t)\eta(p\t)\right]^w}.}  For $p=1$, we recover the familiar counting function $Z_1(\t) = 1/\De(\t)$. More generally, these functions are weight $-w$ modular forms for $\G_0(p)+$ with multiplier system $1/\psi_p(\g)$. Thus the problem of computing the degeneracy of twisted sector DH states has been reduced to computing the Fourier coefficients of a simple class of modular forms.

\subsection{Rademacher Series for $\G_0(p)+$}
\label{sec:CHLrad}
To compute the $\dnp$ exactly, we need an analog of the Rademacher expansion in Eq. \ref{eq:DHrad}. However, the textbook Rademacher series applies only to modular forms for the full modular group. For modular forms of congruence subgroups, care must be taken to appropriately treat each inequivalent cusp of the group. Fortunately, we can avoid this subtlety. The counting functions $\zpt$ extend to modular forms under the AL groups $\gopp$, which have only one cusp. Thus, we can avoid the complications that originate from multiple cusps, and simply construct a single-cusp Rademacher series. This is the approach we take here. However, identical results can also be obtained by directly constructing a two-cusp Rademacher series for the group $\gop$ itself \cite{ethan}.\footnote{After publication, we were made aware of an unpublished thesis \cite{naveen} that also derives Eq. \ref{eq:CHLrad} from the two cusp perspective. This work also presents partial progress towards the results of Section 3, from a complementary perspective in the heterotic frame. We thank N. Prabhakar for bringing this to our attention.} We note that, for the $p=2$ model, similar results were obtained in \cite{0502,0507}.

A detailed and systematic derivation of Rademacher series for all groups commensurate with $\sltz$, and in particular for  congruence subgroups and their extension to AL groups, can be found in Section 3.2 of \cite{Duncan:2009sq}; we will summarize this discussion here. Expository accounts can be found in the review articles \cite{ChengRademacher,Duncan:2014vfa}; a similar analysis for vector-valued modular forms was provided in \cite{2014arXiv1406.0571W}. A general modular form $P^{[m]}_{\G,\psi,w}$ for a congruence subgroup $\G$ of weight $w$ with multiplier system $\psi$ and leading singularity $q^m$ admits a Fourier expansion of the form \eq{P^{[m]}_{\G,\psi,w}(\t) = q^m + \sum_{n\ge0} d_{\G,\psi,w}(n,m)q^n.} For $w\le1$, the Fourier coefficients $d_{\G,\psi,w}$ are given by \eq{d_{\G,\psi,w} = \sum_{\ggg} \exp\left[2\pi i \left(n\frac{d}{c} + m\frac{a}{c} - \frac{w}{4}\right)\right]\psi(\g) \sum_{k\ge0} \left(\frac{2\pi}{c}\right)^{2k+2-w} \frac{(-m)^{k+1-w}n^k}{\G(k+2-w)k!},\label{eq:FDrad}} where we take each representative $\g$ of the cosets $\ggg$ to be of the form $\g=\smallmat{a&b\\c&d}$ and $\psi(\g)$ is the multiplier associated to the transformation $\g$.  

This expansion is extremely general. For our purposes, we can immediately specialize to the case of the $\zpt$, which are weight $-w$ modular forms for $\gopp$ with leading singularity $q^{-1}$. With these assumptions, Eq. \ref{eq:FDrad} simplifies to \eq{\dnp = \sum_{\ggpg} \exp\left[2\pi i\left(n\frac{d}{c}-\frac{a}{c}+\frac{w}{4}\right)\right]\psi_p(\g)^{-1}c^{-2-w}\hi_{1+w}\left[\frac{4\pi}{c}\sqrt{n}\right],\label{eq:dfp}} where $\psi_p(\g)$ is as defined in Eqs. \ref{eq:G0pmult} and \ref{eq:ALmult}.  

It is straightforward to derive Eq. \ref{eq:DHrad} from Eq. \ref{eq:dfp}. For $p=1$, we have $\G_0(1)=\G_0(1)+=\sltz$. The cosets $\G^\infty\backslash\sltz/\G^\infty$ are indexed by positive, relatively prime integers $k>h$. To each such pair, we associate an equivalence class $\ghk$ with canonical representative \eq{\ghk = \left(\begin{array}{cc}h^{-1}&\frac{1}{k}\left(hh^{-1}-1\right)\\k&h\end{array}\right)\label{eq:ghkdh},} where the inverse $h^{-1}$ of $h$ mod $k$ is defined by the condition $hh^{-1} = 1\mod{k}$. Inserting Eq. \ref{eq:ghkdh} into Eq. \ref{eq:dfp}, we find that \eq{d_n^{(1)} = \sum_{k=0}^\infty \sumh k^{-14} \exp\left[2\pi i \left(n\frac{h}{k} - \frac{h^{-1}}{k}\right)\right] \hi_{13}\left[\frac{4\pi}{k}\sqrt{n}\right],} which is exactly Eq. \ref{eq:DHrad}. 

For $p\neq1$, the discussion is similar, but more involved. The equivalence classes $\gpkh$ of $\ggpg$ are again indexed by  coprime integers $k,h$, but now there are two types of equivalence classes, depending on the residue of $k$ modulo $p$. If $k\equiv0\mod{p}$, i.e. if $p$ divides $k$, then $\gpkh$ is of $\gop$ type, and we have \begin{subequations}\label{eq:gkhp} \eq{\gpkh = \left(\begin{array}{cc}h^{-1}&\frac{1}{k}\left(hh^{-1}-1\right)\\k&h\end{array}\right)\label{eq:gkhpa}, \ \ \gcd(k,p)=p.} Conversely, if $\gcd(p,k)=1$, then $\gkhp$ is of AL type, and has form \eq{\gpkh = \frac{1}{\sqrt{p}}\left(\begin{array}{cc}p(ph)^{-1} & \frac{1}{k}\left[(ph)^{-1}ph-1\right] \\ pk & ph\end{array}\right), \ \ \gcd(k,p)=1\label{eq:gkhpb}.} \end{subequations}

We can now plug Eq. \ref{eq:gkhp} into \ref{eq:dfp} to find \seq{\dnp = &\sum_{\substack{k>0\\\gcd(p,k)=p}} \sumh C_p(k,h) \exp\left[2\pi i\left(n\frac{h}{k} - \frac{h^{-1}}{k}\right)\right]\hi_{1+w}\left[\frac{4\pi}{k}\sqrt{n}\right] \\ + &\sum_{\substack{k>0\\\gcd(p,k)=1}} \sumh C_p(k,h) \exp\left[2\pi i\left(n\frac{h}{k} - \frac{(ph)^{-1}}{k}\right)\right] \hi_{1+w}\left[\frac{4\pi}{k}\sqrt{\frac{n}{p}}\right], \numberthis \label{eq:CHLrad}} which is exactly Eq. \ref{eq:IntroRad}. For $p\neq7$, the coefficients $C_p(k,h)$ are defined as \begin{subequations} \eq{C_p(k,h) = \begin{cases}\frac{e^{2\pi iw/4}}{k^{2+w}} & \gcd(k,p) =p \\ \frac{1}{(\sqrt{p}k)^{2+w}} & \gcd(k,p) = 1 \end{cases} } and for $p=7$ we have  \eq{C_p(k,h) = \begin{cases}  \frac{1}{\left(\frac{k}{7}\right)\left(\sqrt{7}k\right)^{5}} & \gcd(k,p) = 1\\ \frac{-i}{(\sqrt{p}k)^5\left(\frac{h}{7}\right)} & \gcd(k,p) = p, \ h \text{ even} \\ \frac{-i}{\left(\sqrt{p}k\right)^5\left(\frac{-7}{h}\right)} & \gcd(k,p)=p, \ h \text{ odd} \end{cases}. } \end{subequations} For $p\neq7$, the $C_p(k,h)$ are independent of $h$, so we can cast Eq. \ref{eq:CHLrad} in a form more reminiscent of Eq. \ref{eq:DHrad}: \eq{d^{(p\neq7)}_n = \sum_{\gcd(p,k)=p} k^{-2-w} \kl(n,-1,k) \hi_{1+w}\left[\frac{4\pi}{k}\sqrt{n}\right] + \sum_{\gcd(p,k)=1}\left(\sqrt{p}k\right)^{-2-w}\kl\left(n,-p^{-1},k\right)\hi_{1+w}\left[\frac{4\pi}{k}\sqrt{\frac{n}{p}}\right]. \label{eq:pneq7rad}}

Eq. \ref{eq:CHLrad} is the main result of this section. The remainder of the paper will be spent providing a gravitational interpretation of this expansion, but the leading term is worthy of special attention before we continue. At large $n$ and for all $p$, the series is dominated by the $k=1$ term, given by \eq{\dnp\Big|_{k=1} = p^{-1-w/2}\hi_{1+w}\left[4\pi\sqrt{\frac{n}{p}}\right].\label{eq:CHLrad1}} Terms of this form are generated by the Fricke involutions $\t\to-1/p\t$. At large $n$, the Bessel functions essentially scale as $\hi_{1+w}\left[\frac{4\pi}{c}\sqrt{n}\right] \sim \exp\left[\frac{4\pi}{c}\sqrt{n}\right].$ As such, this leading term is exponentially larger relative to the successive terms. Moveover, this exponential comes equipped with an entire series of corrections that are polynomial or logarithmic in $n$. This is suggestive of the form of a perturbative expansion; we will see below that this is indeed the correct interpretation of the leading Bessel function in a gravitational context.

\section{The Half-BPS Farey Tale}
\label{sec:macro}

So far all of our discussion has focused on worldsheet aspects of the CHL models. In this section we will discuss the gravitational aspects of these theories, before providing a macroscopic interpretation of the results derived above. In particular, we will compute the degeneracy of black hole microstates in the supergravities dual to the prime-$p$ CHL models, and find detailed agreement with Eq. \ref{eq:CHLrad}; these supergravity theories will be introduced in Section \ref{sec:gravitational}.

After introducing the gravities associated to the CHL models, we will begin to reproduce Eq. \ref{eq:CHLrad}. First, following \cite{0502,0507}, we will in Section \ref{sec:OSV} reproduce the $k=1$ term in the expansion for $\dnp$, which encompases all perturbative corrections to the degeneracy. To obtain nonperturbative corrections to the macroscopic state counts, we will follow a variant of Farey tale \cite{FareyTail,deBoer:2006vg,Manschot:2007ha,Murthy:2009dq} that was used to reproduce the Rademacher expansion for $Z_1(\t) = 1/\De(\t)$ in \cite{Dabholkar:2014ema}. In Section \ref{sec:FT} we will review this approach and explain how to use it to derive Rademacher series for the rest of the $Z_p$.

\subsection{Reduced Rank $\calN=4$ Supergravity}
\label{sec:gravitational}

The CHL models describe compactifications of IIA string theory on (supersymmetry preserving orbifolds of) $K3\times{T}^2$, so at low energies they are well-described by $\calN=4$ supergravity. In addition to the $\calN=4$ supergravity multiplet, which contains a graviton, four spin-3/2 gravitini, six vectors, four spin-1/2 fermions, and a scalar, these theories also contain nontrivial gauge sectors. The total number of spacetime vectors is $r_p$, so we couple the supergravity multiplet to a number $n_V = r_p-6$ of $\calN=4$ vector multiplets. Comparing with Eq. \ref{eq:pRank}, we find \eq{n_V = \frac{48}{p+1} - 2.\label{eq:nvp}} 

Although these are $\calN=4$ theories, it will be useful to consider them instead as $\calN=2$ theories \cite{0409,0502,0507,Dabholkar:2010uh,Dabholkar:2011ec,Gomes:2011zzc,Dabholkar:2014ema}. To do this, we need to decompose the $\calN=4$ multiplets into $\calN=2$ ones; we will follow the decomposition laid out in \cite{0507}. For the vector multiplets, this is simple; each of the $n_V$ $\calN=4$ vector multiplet becomes a $\calN=2$ vector multiplet, plus a hypermultiplet. In what follows, we will mostly ignore the hypermultiplets, as has recently been justified in \cite{Gupta:2012cy}. The supergravity multiplet, similarly, decomposes as a $\calN=2$ supergravity multiplet plus a $\calN=2$ vector multiplet, as well as two $\calN=2$ ``gravitino multiplets", each of which as a spin-3/2 fermion, two vectors, and a spin-1/2 fermion. We will restrict ourselves to states uncharged under the four vectors contained in these two gravitino multiplets \cite{0507}, leaving us with a total of $r_p-4 = n_V+2 = \frac{48}{p+1}$ vectors $A^I$ in the four dimensional theory\footnote{Here and for the rest of the paper, the index $I$ runs between $0$ and $n_V+1$. We will later introduce additional indices $a,b$, which take the $n_V$ values $2,3,\cdots,n_V+1$.}. We will take the $A^I$ to have electric field strengths $e^I$ and couple to electric and magnetic charges $q_I$ and $p^I$, respectively. To each of the $A^I$ we associate a complex scalar moduli field $X^I$; the ratio $X^1/X^0$ is the heterotic axiodilation \cite{0409} on which the $S$-dualities, including the Fricke $S$-duality described in Section 2, act. The $A^I$ and the $X^I$, as well as the metric field $g$, will be the primary variables in the gravity theory. 

As $\calN=2$ theories, the gravities dual to the CHL model are have a prepotential $F^{(p)}(X^I,W^2),$ where here $W^2$ is the graviphoton field strength squared. However, as $\calN=4$ theories, the prepotentials enjoy nonrenormalization theorems. We can therefore write down a quantum-exact prepotential for the $X^I$: \cite{0409,0502,0507,LopesCardoso:1999fsj,Persson:2015jka} \eq{F^{(p)}(X^I,W^2) = -\frac{1}{2}C_{ab}X^aX^b\frac{X^1}{X^0} + \frac{W^2}{128\pi i}\log{Z_p(\tq)},\label{eq:prepotential}} where here $\tq = \exp\left[2\pi i X^1/X^0\right]$. The presence of $Z_p(\t)$ here stems ultimately from a calculation in topological string theory \cite{Bossard:2017wum,Persson:2015jka}. Here, $C_{ab}$ is the intersection matrix of two-cycles of $K3/\Z_p$. 

We will be interested in half-BPS excitations in these theories. In Section \ref{sec:functs} we constructed these states in the heterotic frame as fundamental strings wrapping a circle $S^1\subset T^6$. For macroscopic considerations, it is more convenient to work in a type II picture, where these states are comprised of branes instead of fundamental strings. In the IIA frame, the heterotic states with $n/p$ units of momentum and one unit of winding described in Section \ref{sec:functs} are described by a gas of $n/N$ D0 branes living on the worldvolume of a D4 brane wrapping the (singular) K3 fiber of the compact space.

As discussed above, states in these theories have $r_p$-dimensional electric and magnetic charge vectors $Q$ and $P$, respectively. A general configuration of $q_0$ D0 branes living on $p_1$ D4 branes will have charge vectors \begin{subequations} \eq{Q &=  \left(q_0,p^1,0,\cdots,0\right) \label{eq:qGen} \\ P &= (0,0,\cdots,0) \label{eq:pGen}.} The states we consider have charge vectors of this form, but with \eq{q_0 &= \frac{n}{p} \label{eq:q0} \\ p_1 &= 1 \label{eq:p1}.} \end{subequations} In what follows, we will formally use the generic charge vectors in Eqs. \ref{eq:qGen} and \ref{eq:pGen}, but for twisted sector states it is important to keep in mind the particular assignment of charges in Eqs. \ref{eq:q0} and \ref{eq:p1}.

As with the half-BPS states in the standard compatification of IIA on $K3\times{T}^2$, these states are ``small" black holes, i.e. they have classically vanishing horizon area \cite{Sen:1995in}. After the inclusion of $\a'$ corrections to the low energy action, however, a finite-size horizon appears, and we are therefore free to work in the usual $AdS_2\times{S}^2$ geometry associated to extremal black holes \cite{0409,0502,0507,Dabholkar:2004dq}. 

As usual with extremal black holes in $\calN\ge2$ supergravities, the moduli fields of these black holes are governed by an attractor mechanism. The attractor equations relate the $X^I$ to the $q_I$ and $p^I$\cite{0409,0502,0507,Ferrara:1995ih,Ferrara:1996dd} :

 \begin{subequations}\label{eq:attractorEqs}
     \begin{align}
p^I &= \Re\left[X^I\right] \label{eq:attractorP} \\
q_I &= \Re\left[\d_{X^I} F\left(X^I,W^2=256\right)\right] \label{eq:attractorQ}
     \end{align}
    \end{subequations}
    
Following \cite{0409}, we solve the first attractor equation by writing \eq{X^I = p^I + i\phi^I,} where $\phi^I$ should be thought of as a potential conjugate to $q^I$. Using Eq. \ref{eq:prepotential}, we can solve \ref{eq:attractorQ} to find \begin{subequations}\label{eq:attractorPhi} \begin{align} \phi^0 &= \frac{1}{2}\sqrt{\frac{p^1}{q_0}} \\ \phi^a &= 0. \end{align}\end{subequations} The remaining potential $\phi^1$ is not fixed by the attractor equations. The leading order entropy $S_{BH}$ is given by \cite{0409} \eq{S_{BH} =  -\frac{\pi}{2}C_{ab}\phi^a\phi^b\frac{\phi^0}{p^1} + 4\pi\frac{p^1}{\phi^0} + \pi \phi^0q_0, \label{eq:Sbh}} where the $\phi^I$ are evaluated at the attractor point. Inserting Eq. \ref{eq:attractorPhi}, we find \cite{0409,0502,0507} \eq{S_{BH} = 4\pi\sqrt{q_0p^1} = 4\pi\sqrt{\frac{Q^2}{2}}.\label{eq:SBHanswer}} Thus, the leading-order contribution to the degeneracy $\dnp$ is \eq{\dnp \sim \exp\left(S_{BH}\right) = \exp\left[4\pi\sqrt{\frac{Q^2}{2}}\right].} In light of the definition of $\hi_\n(z)$ in Eq. \ref{eq:bessel} and the charge quantization condition in Eq. \ref{eq:Q2p}, this exactly reproduces the exponential part of the $k=1$ term in Eq. \ref{eq:CHLrad}.

\subsection{Perturbative Corrections}
\label{sec:OSV}
To evaluate the leading order contribution to the entropy, we needed to evaluate Eq. \ref{eq:Sbh} at the attractor point. However, to obtain corrections, we will need to consider fluctuations around the attractor geometry. We will now perform a simple computation to verify the remaining contributions to the $k=1$ term in Eq. \ref{eq:CHLrad}, i.e. Eq. \ref{eq:CHLrad1}. Each of these Bessel functions consists of an infinite tower of perturbative corrections to the degeneracy of black hole microstates, and therefore to the entropy of the black hole. In the spirit of OSV, we can, following \cite{0409,0502,0507}, define a black hole partition function $Z_{BH}$ as a series in the electric potentials $\phi^I$ and magnetic charges $p^I$ by \eq{Z_{BH}\left(\phi^I,p^I\right) = \sum_{q_I} \Omega\left(q_I,p^I\right)e^{-\phi^Iq_I},\label{eq:OSVpartition}} where the sum runs over the allowed lattice of electric charges. The degeneracies $\Omega(q_I,p^I)$ are given by \eq{\Omega\left(q_I,p^I\right) = \left(\frac{1}{2\pi}\right)^{n_V+2} \int d\phi^0d\phi^1 \prod_{a=2}^{n_V+1} d\phi^a\ \exp\left[\F\left(\phi^I,p^I\right) + \phi^0q_0\right],\label{eq:OSV}} where \eq{\F(\phi^I,p^I) \equiv -\pi \Im\left[F^{(p)}\left(p^I+i\phi^I,W^2=256\right)\right]\label{eq:calF}} is the imaginary part of the prepotential $\fp$. Heuristically, $\mathcal{F}$ should be thought of as the free energy of the system. In terms of $\F$, Eq. \ref{eq:Sbh} can be written as \eq{S_{BH} = \F + \pi\phi^0q_0,} again evaluated at the attractor point.

Inserting Eq. \ref{eq:prepotential} into Eq. \ref{eq:calF}, we find that \eq{\F(\phi^I,p^I) = -\frac{\pi}{2}C_{ab}\phi^a\phi^b\frac{p^1}{\phi^0} - \ln\left|\eta(\tq)\eta(\tq^p)\right|^{2w}.}  Plugging into Eq. \ref{eq:OSV}, we have \eq{\Om(q_I,p^I) = \frac{1}{(2\pi)^{n_V+2}} \int d\phi^0\ d\phi^1 \prod_{a=2}^{n_V+1}d\phi^a\ \exp\left[ -\frac{1}{2}C_{ab}\phi^a\phi^b\frac{p^1}{\phi^0} - \ln\left|\eta(\tq)\eta(\tq^p)\right|^{2w} + q_0\phi^0\right].} This integral may be evaluated to yield \cite{0409,0502,0507} \eq{\Om(q_I,p^I) = \frac{1}{\sqrt{\det{C_{ab}}}}\left(p^1\right)^2 \hi_{2+n_V/2}\left[4\pi\sqrt{q_0p^1}\right].\label{eq:OSVanswer}} Up to a prefactor, we therefore have exact matching with the leading Bessel function predicted in the previous section. There is, however, an important caveat. The prefactor $(p^1)^2$ violates duality invariance. This suggests that care must be taken in defining the integration measure $\prod d\phi^I$; this will again be seen in what follows.

\subsection{Nonperturbative Corrections and $AdS_3$ Saddles}
\label{sec:FT}
We saw above how to reproduce perturbative corrections to the degeneracy from macroscopic considerations. In this section we will discuss the interpretation of the nonperturbative corrections, i.e. the $k>1$ terms in Eq. \ref{eq:CHLrad}. The essential logic is that the Rademacher expansion for the worldsheet degeneracies $d_n$ should be dual to the saddle point expansion for the expectation value of a judiciously chosen operator in an $AdS_3$ gravity theory obtained as a lift of the $AdS_2$ near-horizon geometry of the extremal black hole; this is related to the Farey tale \cite{FareyTail,deBoer:2006vg,Manschot:2007ha,Murthy:2009dq}. 

A more rigorous version of this basic argument is provided in the context of AdS$_2$/CFT$_1$ by Sen's quantum entropy function, introduced in \cite{Sen:2008yk,Sen:2008vm,Sen:2009vz}. This formalism interprets the macroscopic degeneracy $\Om(Q^2)$ as the finite part of the expectation value of an Abelian Wilson line inserted on the boundary of $AdS_2$, \eq{\Omega(Q^2) \sim \left<\exp\left[\frac{i}{2}q_I \oint d\th A^I_\th\right]\right>_{AdS_2}^{finite}.\label{eq:pathintegral}} The superscript ``finite" indicates a prescription for removing the IR divergence associated with the infinite volume of $AdS_2$. The expectation value here is defined formally as a path integral over massless supergravity modes; we choose boundary conditions such that the metric is asymptotically $AdS_2$ and matter fields asymptote to their attractor values. In the IIA frame, the (Euclidean) metric, scalars $X^I$, and vectors $A^I$ have attractor values  \cite{Dabholkar:2010uh,Dabholkar:2014ema} \subeqslabel{ds^2 &= v_*\left[\left(r^2-1\right)d\th^2 + \frac{dr^2}{r^2-1}\right] \\ X^I &= X^I_* \\ A^I &= -ie^I_*rd\th,}{eq:attractorvalues} where $v_*$, $X_*^I$, and $e^I_*$ are constants.  

To obtain a gravitational interpretation of the Kloosterman sums associated to the $k>1$ terms, it is necessary to consider an $AdS_3$ lift of the $AdS_2$ near-horizon geometry. More precisely, after lifting the IIA geometry to the M-theory frame, we obtain an $AdS_2\times{S}^1$ geometry, with metric \eq{ds^2 = \left(r^2-1\right)d\th^2 + \frac{dr^2}{r^2-1} + R^2\left[dy - \frac{i}{R}\left(r-1\right)d\th\right]^2,\label{eq:ds2ads2s1}} where $y$ is the coordinate along the M-theory circle and $R$ is the attractor value of its radius. We will consider supersymmetric localization of the three-dimensional path integral around this background given geometry. Full details on the localization calculation are given in \cite{Dabholkar:2010uh,Dabholkar:2011ec,Gomes:2011zzc,Dabholkar:2014ema,Gomes:2013cca,Murthy:2015yfa}, and in particular in \cite{Dabholkar:2014ema} it was shown how to obtain the Fourier coefficients for $Z_1(\t) = 1/\De(\t)$ from this approach. We refer the reader to those works for more details on this construction, and will instead summarize the results of these results.

The path integral localizes on an infinite dimensional set of saddles given by freely acting orbifolds of $AdS_2\times{S}^1$. We orbifold by a rotation of $\frac{2\pi}{k}$ along $AdS_2$, as well as a translation along $S^1$ by $\frac{2\pi h}{k}$. These orbifolds are freely acting so long as $\gcd(h,k)=1$ \cite{FareyTail,deBoer:2006vg,Manschot:2007ha,Murthy:2009dq,Dabholkar:2010uh,Dabholkar:2011ec,Gomes:2011zzc,Dabholkar:2014ema,Maloney:2007ud,Maldacena:1998bw}. We will call each orbifold geometry $M_{k,h}$. The metric on $M_{k,h}$ is given by \eq{ds^2_{k,h} = \left(r^2 - \frac{1}{k^2}\right)d\th^2 + \frac{dr^2}{r^2-\frac{1}{k^2}} + R^2\left[dy - \frac{i}{R}\left(r-\frac{1}{k}\right)d\th + \frac{k}{h}d\th\right]^2, \label{eq:ads2s1OrbifoldMetric}} where now \eq{\th\in\left[0,\frac{2\pi}{c}\right).\label{eq:thDef}} Inserting $k=1,h=0$, this metric can easily be seen to reduce to Eq. \ref{eq:ds2ads2s1}; moreover, asymptotically as $r\to\infty$, these metrics become independent of $k,h$, and hence all of these orbifold geometries contribute to the path integral \cite{Dabholkar:2014ema}. Thus the path integral takes the form \eq{\Om = \sum_k \sumh \Om_{k,h},\label{eq:Zsum}} where by $\Om_{k,h}$ we mean the contribution to $\Om$ from $M_{k,h}$.

To evaluate the $\Om_{k,h}$, it is convenient to realize the $M_{k,h}$ as quotients of $AdS_3$ \cite{Murthy:2009dq,Dabholkar:2014ema,Maldacena:1998bw,Maloney:2007ud}. Topologically, thermal Euclidean $AdS_3$ is a solid two-torus. Its asymptotic boundary is a therefore a $T^2$. Let us choose homology cycles $C_1$ and $C_2$ such that $C_2$ is contractible through the interior of the solid torus. To obtain $AdS_3$ analogs of the orbifolds in Eq. \ref{eq:ads2s1OrbifoldMetric}, one simply chooses new cycles, $\tc_1$ and $\tc_2$, respectively, to be the noncontractible and contractible cycles; in terms of $C_1$ and $C_2$, these are defined as \eq{\left(\begin{array}{c}\tc_1\\\tc_2\end{array}\right) = \cc a&b\\c&d\ccend \left(\begin{array}{c}C_1\\C_2\end{array}\right).\label{eq:ads3orbifold}} Here as usual $\smallmat{a&b\\c&d}$ is an $\sltz$ matrix. By choosing this matrix appropriately, we can obtain an $AdS_3$ version of each orbifold in Eq. \ref{eq:ads2s1OrbifoldMetric}.

In the $AdS^3$ realization of $M_{k,h}$, we have bulk dynamics as well as topological Chern-Simons terms. The bulk dynamics are essentially inherited from the $AdS_2$ framework, and essentially contribute the same Bessel function derived in Section \ref{sec:OSV}, but with the argument scaled so as to reflect the overall volume decrease. The Chern-Simons terms are more subtle. We of course have CS terms from the Abelian gauge fields $A^I$. However, to ensure that the localization supercharge $Q$ remains a symmetry of the orbifold geometries, we must turn on an $SU(2)$ gauge field on the $S^2$, which receives additional CS terms. The CS terms from the $A^I$ and this new $SU(2)$ gauge field will combine with gravitational CS terms to yield precisely the Kloosterman sums in Eq. \ref{eq:CHLrad}.

We will briefly outline how to handle each of these terms in the case of $Z_1(\t)=1/{\De(\t)}$; more detail can be found in the original works \cite{Dabholkar:2010uh,Dabholkar:2011ec,Gomes:2011zzc,Dabholkar:2014ema,Gomes:2013cca}. After localization, the AdS$_3$ path integral is given as a sum over saddle points, as indicated in Eq. \ref{eq:Zsum}. To evaluate the contribution $\Om_{k,h}$ of each saddle, one simply has to perform a finite dimensional integral over the moduli fields $X^I$, which for convenience are usually repackaged into the $\phi^I$, as in Section \ref{sec:OSV}. As alluded to above, the evaluation of the $\Om_{k,h}$ roughly speaking splits into two parts, one coming from the bulk dynamics and the other coming from the topological terms. To obtain the contribution from the dynamical terms, one must solve the M-theory frame attractor equations, which give the $AdS_2\times{S}^1\times{S}^2$ equations of motion for the moduli fields. It was shown in \cite{Gomes:2013cca} that the five-dimensional attractor equations reduce exactly to the four-dimensional ones discussed in Section \ref{sec:OSV}, and that therefore the dynamical portion of the path integral, for the leading saddle at least, is given by Eq. \ref{eq:OSVanswer}.

The subleading saddles are constructed from freely acting orbifolds of $AdS_3$. Therefore, these orbifolds cannot effect the dynamics, and the dynamical contribution is of the same form given in Eq. \ref{eq:OSVanswer}. However, the global volume of the spacetime of these orbifolds is reduced, so the argument of the Bessel function in Eq. \ref{eq:OSVanswer} must be rescaled to reflect the decrease in volume. The appropriate scaling for the orbifold generated by an arbitrary $\sltz$ matrix $\g=\abcd$ can be shown to be \eq{\Om\left[\g\right]\Big|_{\text{Bessel}} \sim \hi_{2+n_V/2}\left[\frac{4\pi}{k}\sqrt{\frac{Q^2}{2}}\right]. \label{eq:Zbessel}} In this equation we have omitted the prefactor, essentially because the integration measure after localization is poorly understood. This leaves an ambiguity in the overall prefactor of the expression, which will be discussed below.

The topological terms are more subtle, but can be understood through brute force calculations. There are two varieties of Chern-Simons terms. The first simply corresponds to CS terms for the U(1) gauge fields $A^I$ living in the vector multiplets. Supersymmetry considerations indicate that the Abelian gauge fields must be flat at infinity; by considering the holonomies of flat gauge fields about the contractible and noncontractible cycles, it is straightforward to show that, after the addition of appropriate boundary terms, the Abelian CS terms yield \cite{Dabholkar:2014ema,Murthy:2009dq} \eq{Z[\g]\Big|_{\text{Abelian CS}} \sim \exp\left(2\pi i \frac{Q^2}{2} \frac{d}{c}\right).} In addition, in order to ensure that the localizing supersymmetry remains a symmetry of the higher saddles, one must turn on an SU(2) gauge field on the $S^2$ portion of the metric \cite{Dabholkar:2014ema}. These gauge fields also have CS terms, which can be shown to give a contribution of the form \eq{Z[\g]\Big|_{\text{Nonabelian CS}} \sim \exp\left(-2\pi i \frac{a}{c}\right).}

Putting it all together, we have that the contribution $\Om[\g]$ to the gravitational path integral from a saddle point constructed from $AdS_3$ by an orbifold matrix $\g=\smallmat{a&b\\c&d}$ is given by  \cite{Dabholkar:2014ema} \eq{\Om[\g] &= \Om\left[\g\right]\Big|_{\text{Bessel}} Z[\g]\Big|_{\text{Abelian CS}}  Z[\g]\Big|_{\text{Nonabelian CS}} \sim \exp\left[2\pi i\left(\frac{Q^2}{2}\frac{d}{c} - \frac{a}{c}\right)\right] \hi_{13}\left[\frac{4\pi}{c}\sqrt{\frac{Q^2}{2}}\right].\label{eq:og}} This is the contribution from each saddle, and now we must sum over the different saddles. For the $p=1$ model, i.e. the unorbifolded compactification, the saddle points are simply indexed by elements $\gkh$ of $\G^\infty\backslash\sltz/\G^\infty$, which we can simply take to have the form given in Eq. \ref{eq:ghkdh}, i.e. we take $\gkh = \smallmat{h^{-1}&\\k&h}$. The leading term saddle point is generated by the $S$-transformation matrix $g_{1,0} = \smallmat{0&-1\\1&0}$. Inserting this matrix into Eq. \ref{eq:og}, we find that \eq{\Om\left[\g_{1,0}\right] \sim \hi_{13}\left[4\pi\sqrt{\frac{Q^2}{2}}\right].} By construction, this agrees exactly with Eq. \ref{eq:OSVanswer}. However, given that $Q^2/2 = n$ in this model, it also agrees exactly with the leading term in Eq. \ref{eq:DHrad}. For more general $\gkh$, we have \eq{\Om\left[\gkh\right] \sim  \exp\left[2\pi i\left(\frac{Q^2}{2}\frac{h}{k} - \frac{h^{-1}}{k}\right)\right] \hi_{13}\left[\frac{4\pi}{k}\sqrt{\frac{Q^2}{2}}\right].} We can therefore straightforwardly sum over the remaining $\gkh$ to find that \eq{\Om \equiv \sum_{k,h}\Om\left[\gkh\right] \sim \sum_{k=1}^\infty \sumh \exp\left[2\pi i \left(\frac{Q^2}{2}\frac{h}{k} - \frac{h^{-1}}{k}\right)\right]\hi_{13}\left[\frac{4\pi}{k}\sqrt{\frac{Q^2}{2}}\right].} Apart from the factor of $k^{-14}$ in each term, this agrees exactly with the Rademacher expansion for the Fourier coefficients $d_n$ of $1/\De(\t)$ given in Eq. \ref{eq:DHrad}. How to obtain this factor is an open question, and involves properly deriving the integration measure on each localizing saddle \cite{Gomes:2017bpi,Gomes:2017eac,Gomes:2015xcf,Dabholkar:2014ema}. We will not attempt to solve this issue in this work.

We would now like to generalize the above discussion to the remainder of the prime-$p$ CHL models. For the $p\neq1$ models, we will need to make two modifications to the above argument. Both the set of saddles to sum over, and the contribution of each saddle, will be modified. We will begin by discussing the appropriate saddle points for the $p\neq1$ models. The saddle point expansion in twisted gravitational theories have been discussed in \cite{Duncan:2009sq}, where it is argued that the appropriate saddle points in the gravity theory dual to $g$-twisted sector of a two-dimensional CFT are given by elements of $\G^\infty\backslash \G_g/\G^\infty,$ where $\G_g$ is the group of modular transformations preserving the $g$-twist. Although this argument was originally presented in the context of monstrous moonshine, from a gravitational perspective it is essentially rooted in the topology of solid tori, and so we expect this argument to carry over here. In our case, the $\G_g$ are simply $\G_0(p)+$.We therefore have two varieties of saddles: those corresponding to $\G_0(p)$ transformations, which are inherited directly from the parent theory discussed above, and a second class corresponding to Atkin-Lehner type transformations. These saddles are all constructed from orbifolds of $AdS_3$ in the usual way, but now the appropriate matrices $\g$ are of the form given in Eq. \ref{eq:gkhp}. 

This poses a minor puzzle. To obtain the leading contribution to the macroscopic entropy in the $p=1$ model, we considered the saddle point obtained by orbifolding by the $S$-transformation $\t\to-1/\t$. However, for $p\neq1$, this is not an element of $\G^\infty\backslash\gopp/\G^\infty$, and hence this saddle does not appear in the orbifolded theories. It is therefore natural to wonder which saddle point gives the dominant contribution to the gravitational path integral. The natural analog of the $S$-transformation in $\gopp$ is the Fricke involution, $\t\to-1/p\t$, and already from Eq. \ref{eq:gkhp} we can read off that this saddle point will indeed be the dominant contribution to the macroscopic calculation.

We now move on to the needed modifications to Eq. \ref{eq:og}, of which there are again two. The first is simply to reflect that the CHL models have fewer vector fields than does the usual compactification of IIA on $K3\times{T}^2$. This simply decreases the index of the Bessel function in Eq. \ref{eq:og}, as is already written in Eq. \ref{eq:Zbessel}. The other modification is more subtle; the charge $Q^2$ of the black hole itself must be rescaled.

To see why, it is easiest to proceed directly to the saddle point expansion of $\Om$. We have argued above that the leading saddle should correspond to the Fricke involution, i.e. the $\sltr$ matrix $\g_{1,0}^{(p)} = \frac{1}{\sqrt{p}}\smallmat{0&-1\\p&0}$. If we take Eq. \ref{eq:og} and simply modify the index of the Bessel function, we find that \eq{\Om\left[\g_{1,0}^{(p)}\right] \sim \hi_{2+n_V/2}\left[\frac{4\pi}{\sqrt{p}}\sqrt{\frac{Q^2}{2}}\right].} Inserting Eqs. \ref{eq:nvp} and \ref{eq:Q2p}, this becomes \eq{\Om\left[\g^{(p)}_{1,0}\right] \sim \hi_{1+w}\left[\frac{4\pi}{p}\sqrt{n}\right].} In addition to the usual prefactor, this only agrees with \ref{eq:CHLrad1} up to a factor of $\sqrt{p}$. This suggests that the charge $Q^2/2$ must be rescaled as \eq{\frac{Q^2}{2} \in \frac{1}{p}\Z \to\frac{pQ^2}{2}\in\Z \label{eq:rescaling}} so that the appropriate modification of Eq. \ref{eq:og} for the $p\neq1$ models should be \eq{\Om[\g] = \exp\left[2\pi i \left(\frac{pQ^2}{2}\frac{d}{c} - \frac{a}{c}\right)\right] \hi_{2+n_V/2}\left[\frac{4\pi}{k}\sqrt{\frac{pQ^2}{2}}\right],\label{eq:Omgp}} where as before we use the placeholder $\g = \smallmat{a&b\\c&d}$. 

We do not have a first-principles derivation of this rescaling. The need for similar scalings has been previously observed in the literature, in Eq. 4.61 of \cite{Gomes:2017zgz} and the following discussion, so this is not unprecedented. It is, however, rather dissatisfying. Essentially, we need to rescale the charges of the $AdS_3\times S^2$ geometry relative to those of the $AdS_2\times S^2$ geometry; the rescaling should not be thought of as living in the IIA frame, since the fractional grading is necessary to ensure matching of Eq. \ref{eq:OSVanswer} to Eq. \ref{eq:CHLrad}. The foremost reason why the M-theory lift of the IIA geometry is needed is the presence of the Chern-Simons terms, which contribute the Kloosterman sums. It is therefore quite reasonable to suspect that the correct explanation of the rescaling comes from a careful analysis of the Chern-Simons terms and their boundary conditions\footnote{This was suggested to us by J. Gomes.}. An immediate consequence of the fractional charge grading is that the Chern-Simons invariants are themselves fractional, regardless of whether we compute them in a $\gop$ saddle or an AL saddle; this can be seen directly from the results of Section 4 of \cite{Dabholkar:2014ema}, where these quantities are computed in a very general framework. The rescaling in Eq. \ref{eq:rescaling} would immediately remedy this, and restore integrality of the Chern-Simons invariants. Although it is possible that making this argument rigorous would lead to a derivation of Eq. \ref{eq:rescaling}, and therefore of Eq. \ref{eq:Omgp}, we will abandon this line of inquiry for now, and simply assume Eq. \ref{eq:Omgp}.

From Eq. \ref{eq:Omgp}, it is straightforward to derive a Rademacher-type expansion. We have \eq{\Om = \sum_{k=0}^\infty \sumh \Om\left[\gkhp\right].} Inserting Eqs. \ref{eq:gkhp}, \ref{eq:Q2p}, and \ref{eq:Omgp}, this becomes \eq{\Om^{(p)} = \sum_{\substack{k > 0\\\gcd(k,p)=p}} \kl(n,-1,k)\hi_{2+n_V/2}\left[\frac{4\pi}{k}\sqrt{n}\right] + \sum_{\substack{k > 0\\\gcd(k,p)=1}} \kl(n,-p^{-1},k)\hi_{2+n_V/2} \left[\frac{4\pi}{k} \sqrt{\frac{n}{p}}\right].} As before, apart from the coefficients $C_p(k,h)$, this reproduces Eq. \ref{eq:CHLrad} exactly, but now we have the lingering question of the charge rescaling.

\section{Conclusion and Outlook}
\label{sec:conclusion}
In this paper we have considered twisted-sector half-BPS states in toroidally compactified heterotic string theory. These states are counted by modular forms $Z_p(\t)$ which transform covariantly under discrete subgroups of $\sltr$ known as Atkin-Lehner groups. By considering these subgroups, we derived a Rademacher expansion for the degeneracy of these BPS states in Eq. \ref{eq:CHLrad}. We then compared this to a macroscopic calculation using the framework of supersymmetric localization, and found partial matching, up to a prefactor that requires more precise determination of the integration measure and a rescaling of the bulk charge. We have thus made progress towards a complete understanding of the twisted sector spectra of these theories from two perspectives.  

We will close by suggesting avenues for further study. The macroscopic calculation presented in Section \ref{sec:macro} stem ultimately from the program of ``exact holography," discussed in e.g. \cite{Dabholkar:2010uh,Dabholkar:2011ec,Gomes:2011zzc}. There are further interesting routes that could be taken to advance the study of CHL models in this framework. One obvious further question is to explore the degeneracies of quarter-BPS states, whose counting functions for these states are known \cite{JatkarSen,Paquette:2017gmb,Cheng:2016org}, in these models in a similar framework. Gravitational investigations of similarly twisted the quarter-BPS counting functions were presented in e.g. \cite{Sen:2009vz,Sen:2009md,Sen:2010ts,Chowdhury:2014lza}, but none of these studies employ the localization techniques used here. Indeed, even the untwisted quarter-BPS index is not completely understood from this perspective; see \cite{Ferrari:2017msn,Gomes:2017eac,Gomes:2017bpi,Gomes:2017zgz} for progress towards deriving and interpreting a Rademacher-type expansion for the quarter-BPS spectrum.

In a related vein, it would be interesting to analyze the remaining CHL models. Although the classification of CHL models has recently been completed \cite{Gaberdiel:2011fg,Paquette:2017gmb,Persson:2015jka,Cheng:2016org}, here we have only considered the simplest CHL models. It would be interesting to extend these results to the full list of CHL models. A priori, there is no reason this cannot be done, although for the majority of the CHL models the gravitational description remains unexplored. One immediate technical hurdle is that the analogues of $\G_0(p)+$ in the CHL models we have not considered are much more complicated than in the models we have discussed. Even with this hurdle, the results of \cite{Duncan:2009sq,ethan} should enable the construction of Rademacher series for the half-BPS counting functions.

We will finish with one more speculative proposal. As twisted sector counting functions, the $Z_p$ play a similar role to the McKay-Thompson series of monstrous moonshine\footnote{For a comprehensive introduction, see e.g. \cite{Duncan:2014vfa}.}. Indeed, the argument used in Section \ref{sec:CHLrad} can be modified slightly to derive a Rademacher expansion for the Fourier coefficients of the McKay-Thompson series instead of the CHL counting functions \cite{2015arXiv150803742L}. This was the original motivation behind the results of \cite{Duncan:2009sq}. A key feature of moonshine is its ``genus zero property," i.e. that each of the McKay-Thompson series is the hauptmodul of a discrete, genus-zero subgroup of $\sltr$. Until recently, the physical interpretation of the genus-zero property remained unclear, but recent work \cite{Paquette:2016xoo,Paquette:2017xui} has shed some light on its origins.

In \cite{Duncan:2009sq}, it was argued that the genus zero property is equivalent to the fact that the McKay-Thompson series can be obtained from a Rademacher sum for the appropriate group. It was further suggested that this could be explained directly by constructing the McKay-Thompson series explicitly as counting functions for twisted theories of $AdS_3$ quantum gravity, as then the saddle point expansion in the bulk would give the desired Rademacher series. This was motivated by the proposal in \cite{Witten:2007kt} that the minimal theory of chiral gravity in $AdS_3$ should be dual to the monster CFT, and therefore naturally admit an action of the monster group.

Although aesthetically pleasing, there have been issues making this construction rigorous, essentially because, at $c=24$, the monster module is not dual to a weakly curved bulk gravity theory, which requires large central charge. However, here we have argued that the use of supersymmetric localization allows precise construction of twisted sector bulk counting functions. Therefore, if the $j$ function was constructed as an untwisted counting function of some supergravity theory, then the arguments here could be straightforwardly generalized to construct the McKay-Thompson series.  

The major obstruction to constructing the McKay-Thompson series in this way is the existence of an appropriate gravity theory. A natural first conjecture is that such a theory can be found by replacing the left-movers on the heterotic worldsheet with the usual moonshine module; this is essentially the model considered in \cite{Paquette:2016xoo,Paquette:2017xui}. The usual Dabholkar-Harvey counting argument would then suggest that a class of half-BPS states are enumerated by the $j$ function, in exactly the same way that the half-BPS states in the usual compactification of heterotic string theory on $T^6$ are counted by $1/\De(\t)$. Even though this theory is still at $c=24$, and therefore not dual to weakly curved gravity, it is tempting to conjecture that a localization approach like the one employed here could enable us to  obtain a macroscopic perspective on the Rademacher expansion for the Fourier coefficients of $j(\t)$, and from there the McKay-Thompson series. However, this theory is a compactification to two dimensions instead of four, and therefore, as argued in \cite{Zimet:2018dev}, it is unclear how to think of its bulk, and therefore it is not obvious how the arguments of Section \ref{sec:macro} can be applied.

\section*{Acknowledgments}
\addcontentsline{toc}{section}{Acknowledgments}
We thank N. Benjamin, S. Kachru, N. Paquette, B. Rayhaun, and M. Zimet for extremely helpful discussions, and especially for comments on drafts of this paper. We are also grateful to E. Sussman for collaboration on early stages of this work, to J. Gomes for useful discussions, and to J. Duncan for very helpful comments on a final draft.  R.N. is funded by the Stanford University Physics Department, a Stanford University Enhancing Diversity in Graduate Eduction (EDGE) grant, and by NSF Fellowship number DGE-1656518. 

\addcontentsline{toc}{section}{References}
\bibliographystyle{JHEP}
\bibliography{RademacherBib}

\end{document}